\documentclass[journal,twoside,web]{ieeecolor}
\usepackage{tmi}
\usepackage[style=ieee,doi=false,isbn=false,url=false,eprint=false]{biblatex}
\usepackage{amsmath,amssymb,amsfonts}
\usepackage{algorithmic}
\usepackage{graphicx}
\usepackage{textcomp}

\usepackage{bbm}
\usepackage{booktabs}
\usepackage{mathtools}
\DeclareMathOperator*{\argmax}{arg\,max}

\def\BibTeX{{\rm B\kern-.05em{\sc i\kern-.025em b}\kern-.08em
    T\kern-.1667em\lower.7ex\hbox{E}\kern-.125emX}}
\begin{document}
\title{LapGM: A Multisequence MR Bias Correction and Normalization Model}
\author{Luciano Vinas*, Arash A. Amini, Jade Fischer, and Atchar Sudhyadhom
\thanks{Manuscript received; This work was was partially supported by NIBIB of the National Institutes of Health under Award R21EB026086. \textit{Asterisk indicates corresponding author}.}
\thanks{*L. Vinas and A. A. Amini are with the Department of Statistics, University of California, Los Angeles, CA 90095 USA (email: lucianovinas@g.ucla.edu; aaamini@ucla.edu).}
\thanks{J. Fischer is with the Department of Medical Physics, University of Victoria, Calgary, AB, Canada (email: jadefischer@uvic.ca).}
\thanks{A. Sudhyadhom is with the Department of Radiation Oncology at Dana-Faber Cancer Institute and the Brigham and Women's Hospital, Harvard Medical School, Boston, MA 02215 (email: ASudhyadhom@bwh.harvard.edu).}}
\maketitle

\begin{abstract}
A spatially regularized Gaussian mixture model, LapGM, is proposed for the bias field correction and magnetic resonance normalization problem. The proposed spatial regularizer gives practitioners fine-tuned control between balancing bias field removal and preserving image contrast preservation for multi-sequence, magnetic resonance images. The fitted Gaussian parameters of LapGM serve as control values which can be used to normalize image intensities across different patient scans. LapGM is compared to well-known debiasing algorithm N4ITK in both the single and multi-sequence setting. As a normalization procedure, LapGM is compared to known techniques such as: max normalization, Z-score normalization, and a water-masked region-of-interest normalization. Lastly a CUDA-accelerated Python package \texttt{lapgm} is provided from the authors for use.
\end{abstract}


\section{Introduction}
\label{sec:introduction}
\IEEEPARstart{T}{he} maturation of machine learning methods has brought significant performance improvements to previously difficult image analyses tasks such as image segmentation \cite{Ronneberger15}, anomaly detection \cite{Dvorak13}, and modality translation \cite{Yi19}. That is not to say the gains realized by these methods are not without their own set of difficulties. In practice, large quantities of quality data are needed first before a performance ceiling is reached. Particularly in the case of medical research, it has been showed that training on poor-quality datasets leads to generalization issues such as out-of-distribution errors \cite{Jung20} and artifact generation \cite{Liang19,Vinas21}.

In order to minimize such errors, it is worthwhile to consider algorithmic alternatives which may be used to improve existing datasets through data-cleaning. Ideally, this data-cleaning routine should utilize domain-specific knowledge and be independent of any initial, training data composition. Following this principle, we show how the unsupervised method of gradient-regularized Gaussian mixtures can be used to correct for strong intensity inhomogeneity artifacts found in multi-coil parallel magnetic resonance imaging (MRI) reconstructions.

MRI intensity inhomogeneities artifacts, also known as bias fields, are non-anatomical intensity variations that vary slowly in the reconstructed image space \cite{Belaroussi06}. We will refer to the prominence of a bias field as a bias field's strength. The strength of a bias field depends both on patient geometry and radiofrequency (RF) receiver coils positioning in a magnetic resonance (MR) scanner \cite{Asher10}. Completely correcting for individual coil contributions at scan time can be difficult and any undercorrection may lead to the appearance of bias fields. These artifacts are made more prominent in multi-coil setups, where multiple coils interfere to produce stronger bias fields.

For MR images with strong bias fields, we found that state-of-the-art bias corrections \cite{Tustison10,Dong14} would either under correct the bias field, leaving the image artifact intact, or over correct the bias field, reducing tissue contrast in the image. Following the Bayesian perspective of log-bias field generation suggested by W. Wells \cite{Wells96}, we highlight a class of covariance matrices that satisfy the low-pass property sought by W. Wells while remaining flexible enough to correct for strongly biased MR images. This class of matrices with sparse inverses are customizable under different weight transformations which allows for efficient and targeted correction on MR images.

Additionally, we use the fitted parameters of the Bayesian mixture model to perform a MR intensity normalization on different groups of patient scans. By combining debiasing and normalization in one step, our method decreases the possibility of chaining post-processing artifacts in the MR data cleaning pipeline. The end impact of which is a streamlined process allowing for consistent and interpretable learning on MR sampled data.

\section{Methods}\label{sec:methods}
Begin by considering an MR image that, due to improper coil sensitivity correction, has been corrupted by a spatial, multiplicative gain field. This slow-varying multiplicative field will be the bias field of our image. Our analysis will include the case that the MR scanner in question may be able to perform multiple imaging sequences at once. In the case of multiple sequences, we assume the bias field remains relatively constant between the different sequences and image reconstruction effects.

More clearly, for an $m$ sequence scan with $n$ voxels per scan let $X = [X_1,\ldots, X_n]\in\mathbb{R}^{m\times n}$ be the matrix of log-intensity measurements corrupted by some additive log-bias field $B\in\mathbb{R}^n$. Let $Z_i\in [K]$ be the tissue group label for voxel $i$. The multi-sequence, log-intensity vector $X_i$ will be conditionally characterized by the multivariate normal
$$X_i\,|\,(Z_i =k,B_i)\sim\mathcal{N}(\mu_k + B_i\mathbbm{1}_m,\Sigma_k),$$
where $\mathbbm{1}_m$ is the $m$-dimensional ones vector and $\mu_k\in\mathbb{R}^m$, $\Sigma_k\in\mathbb{R}^{m\times m}$ are unknown Gaussian parameters. For simplicity, a basic categorical prior is assumed on $Z_i \sim \text{Cat}(\pi)$ where tissue group $k$ has independent proability $\pi_k$ of appearing. Additionally we will assume all Gaussian parameters $\theta = \big(\pi, \{\mu_k\}_{k=1}^K, \{\Sigma_k\}_{k=1}^K\big)$ are pulled from an improper uniform prior distribution. Next incorporating knowledge that $B$ varies slowly in space, we consider the following Gaussian prior
$$B\sim \mathcal{N}\big(0,\tau L^\dagger\big),$$
where $L^{\dagger}$ is the pseudo-inverse of the graph Laplacian associated with spatial structure of our scan and parameter $\tau^{-1}$ is the regularization strength of the log-bias gradient penalty. The precision matrix $L$ will be explained in more detail at Section \ref{sec:lap_wgt}.

We will consider a posterior probability maximization for our model optimization
\begin{equation}\label{eq:lap_gm_obj}
\max_{\theta,B} p_\theta\big(B\,\big|\,\{X_i\}_{i=1}^n\big).
\end{equation}

\begin{figure*}[t]
\centering
\includegraphics[width=12.5cm]{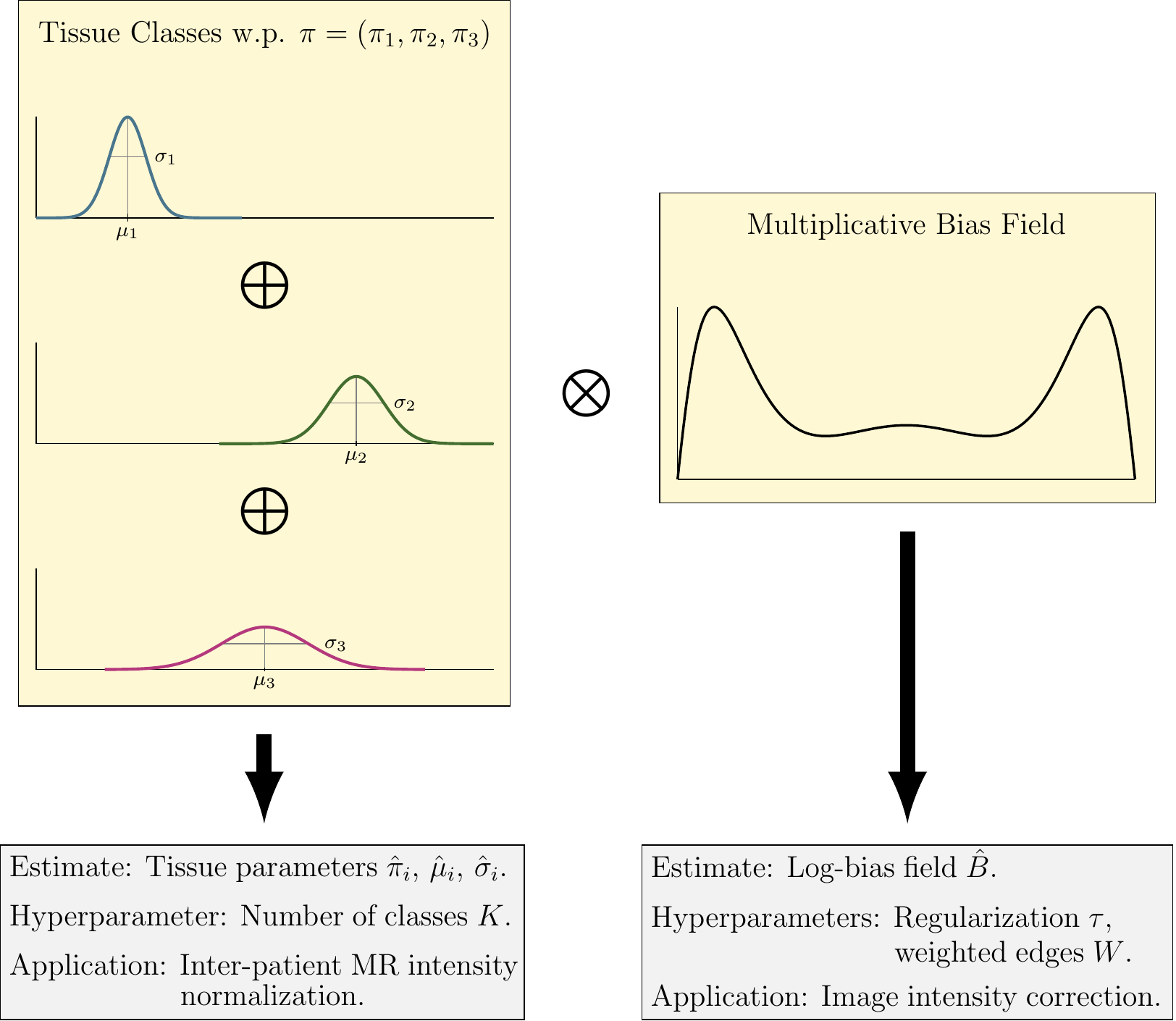}
\caption{Model diagram of LapGM and its use cases. Model assumes intensity distribution follows a Gaussian mixture with additional spatial information determined by a bias field.}
\end{figure*}

In Appendix \ref{app:A1} we show how the minorization-maximization view of expectation-maximization (EM) \cite{Wu10} can be applied to (\ref{eq:lap_gm_obj}) by iteratively optimizing
\begin{equation}\label{eq:Q}
Q\big(\theta,B\,\big|\, \theta^{(t)},B^{(t)}\big) \coloneqq \mathbb{E}_{p_{\theta^{(t)}}(Z\,|\,X,B^{(t)})}[\log p_\theta(B,X,Z)],
\end{equation}
with
\begin{equation}\label{eq:Q2}
\big(\theta^{(t+1)},B^{(t+1)}\big) = \argmax_{\theta,B} Q\big(\theta,B\,\big|\, \theta^{(t)},B^{(t)}\big).
\end{equation}
Note that in the optimization's current state, the parameters $\theta^{(t+1)}$ and $B^{(t+1)}$ cannot be optimized independently of each other. Instead we decouple the expectation step from Gaussian and bias field estimation to produce the following couple closed-form updates:
\begin{itemize}
\item Expectation update
\begin{equation}\label{eq:expect}
w_{ik}^+ = \frac{\pi_k\, p_{\theta}\big(X_i\,|\,Z_i=k,B\big)}{\sum_{\ell=1}^K\pi_l\, p_{\theta}\big(X_i\,|\,Z_i=\ell,B\big)}.
\end{equation}

\item Gaussian update
\begin{align}
\pi_k^+ &= \frac{1}{n}\sum_{i=1}^n w_{ik},\label{eq:cat}\\
\mu_k^+ &= \frac{1}{\sum_{i=1}^n w_{ik}}\sum_{i=1}^n w_{ik}(X_i - B_i\mathbbm{1}_m)\label{eq:mean},\\
\Sigma_k^+ &= \frac{1}{\sum_{i=1}^n w_{ik}} \sum_{i=1}^n w_{ik}(X_i - \mu_k^+ - B_i\mathbbm{1}_m)\nonumber\\
&\hspace{3cm}\cdot(X_i - \mu_k^+ - B_i\mathbbm{1}_m)^\top.\label{eq:cov}
\end{align}

\item Bias field update
\begin{align}
B^+ &= \Big(\frac{1}{\tau}L + \sum_{k=1}^K\mathbbm{1}_m^\top \Sigma_k^{-1}\mathbbm{1}_m\, \text{diag}(w_{\cdot k})\Big)^{-1}\nonumber\\
&\;\cdot\Big(\sum_{k=1}^K \text{diag}(w_{\cdot k})(X-\mu_k\mathbbm{1}_n)^\top\Sigma_k^{-1}\mathbbm{1}_m\Big).\label{eq:bias}
\end{align}
\end{itemize}
Notation $(\cdot)^+$ indicates a subsequent iteration parameter estimate given the current parameter estimate $(\cdot)$. These block updates may be carried out in a randomly-permuted order \cite{Nesterov12,Sun20} for increased parameter exploration. The updates are derived using the first order conditions of a concave objective. Derivations can be found in Appendix \ref{app:A2}.

\subsection{Connection to Previous Works}

The Laplacian-regularized, Gaussian mixture can be softly understood in the general framework of W. Wells \cite{Wells96} where Gaussian prior 
$$B \sim \mathcal{N}(0,\tau \, \psi_B)$$
uses a low-pass filter matrix $\psi_B$ for its covariance. To understand this frameworks connection to (\ref{eq:expect}-\ref{eq:bias}), note that the Laplacian matrix $L$ picks out high frequency content, such as edges, from areas of changing contrast. Under this frequency view, the inverse process $L^\dagger$ can be seen as a low-pass transformation which degrade high frequency content with extraneous low frequency content. As an added benefit, $L^\dagger$ can be understood as a proper inverse to signals $B$ with zero-mean.

This formulation, although exceedingly general, faced a combination of design and computational issues. The first difficulty was to construct a matrix $\psi_B$ which was positive-semidefinite and shared similarities to a low-pass filter. The second difficulty grappled with the computational cost of inverting
$$H = \frac{1}{\tau}\psi_B^{-1}+\sum_{k=1}^K\mathbbm{1}_m^\top (\Sigma_k^{-})^{-1}\mathbbm{1}_m\, \text{diag}\big(w_{\cdot k}^-\big),$$
which itself contained a matrix inverse $\psi_B^{-1}$.

The first issue could be addressed fairly generally by considering positive-definite kernel $K$ with fixed window $T$ and expanding $K$ to its Toeplitz matrix form $\psi_B$ for image dimensions $[n]^d$. This procedure would produce large but sparse covariance matrices with non-zero entries on the order $\mathcal{O}(n T^d)$. The downside to this approach is that there is no guaratee $\psi_B^{-1}$ itself will be sparse and in practice a dense inverse seems to be common. A dense $\psi_B^{-1}$ would make $H$-inversion computationally intractable in an iterative EM method. This problem may be partially solved by computing a convex program on $\psi_B$ which produces approximate sparse inverses $\hat{\psi}_B^{-1}$ depending on some known sparseness value $\alpha$ \cite{Friedman07}. However it is not clear how the forced sparseness of $\hat{\psi}_B^{-1}$ with parameter $\alpha$ will affect the low-pass properties of $\hat{\psi}_B$, potentially compromising the effectiveness and validity of the low-pass prior.

In the original paper of W. Wells, this was circumvented by assuming $H^{-1}$ could be approximated by a uniform filter matrix with kernel window of around 15 to 30 pixels. While computationally tractable, this heuristic is not directly connected to any known covariance matrix and, for its current setting, the large smoothing window could affect correction effectiveness in the case of compact and prominent inhomogeneities. 

By beginning with a penalty $B^\top L B$ and working backward to a generative model, we are able to avoid the inversion process $L^\dagger \to L$, enforce sparsity for $H$, and maintain the low-pass intuition for covariance matrix $L^\dagger$.

\subsection{Gradient Weighting Heuristic}\label{sec:lap_wgt}

It is common in multi-coil setups for bias intensity to increase towards the boundary of the patient anatomy. In order to account for the spatial dependence in multi-coil bias fields, consider the graph Laplacian of an undirected graph $G = (V,E)$ with non-negative edge-weights $W$
$$L_{ij} = \begin{cases}-W_{ij},&\text{if } i\neq j,\\
\sum_{k \in \text{Nbr}(i)} W_{ik},&\text{if } i = j,\end{cases} $$
where $\text{Nbr}(i) \coloneqq \{j\in[n]:\, (i,j) \in E\;\text{or}\; (j,i) \in E\}$. A reasonable heurestic would be to more sharply relax the gradient penalty as voxels approach the boundary of the patient's anatomy. One reasonable choice could be the inverse power function
$$h_\alpha(x,y,z) = (x^2+y^2)^{-\alpha},\quad \text{with } \alpha \geq 0,$$
with cylindrical symmetry about some $z$-axis. For efficiency, the Laplacian edge-weights are constructed through a set of vertex evaluations
$$W_{ij} = \textbf{1}\{i > j\}\cdot h(V_j) + \textbf{1}\{i < j\}\cdot h(V_i),$$
where $V_i$ is the spatial position of the $i$th voxel. For a $d$-dimension grid graph, distinct indices $i,j\in[n]^d$ are ordered as $i > j$ if and only if there exists some $k^\prime \in [d]$ such that
$$\forall k\geq k^\prime,\,i_k > j_k\quad\text{and}\quad\forall k < k^\prime,\, i_k = j_k.$$
This inequality evaluation can be modified independently from the rest of the Bayesian model.

\section{Materials and Experiments}

We compare LapGM to the industry-standard N4ITK (N4) debiasing method. All calculations were run on a 64-core Linux machine equipped with a 3090 Nvidia GPU. The LapGM model was run using the author provided CUDA-accelerated Python package \texttt{lapgm}\footnote{Code available at https://github.com/lucianoAvinas/lapgm.}, while N4 was run using SimpleITK's \cite{Lowekamp13} multi-threaded CPU implementation.

LapGM and N4 were compared using simulated and real-world data. Performance on simualted data was evaluated using a 50-50 validation-testing split. During the validation phase a total of 120 hyperparameter combinations were evaluated for both LapGM and SimpleITK's N4. As suggested by the original N4 authors, all images were downsampled using a 2-factor downsample before running estimation. At inference all estimated bias fields were then upsample to their original image dimensions.

Simulated data was generated using bias simulation software \texttt{biasgen} \cite{Vinas22} in conjuction with the noiseless, 0\% RF non-uniformity BrainWeb dataset \cite{Cocosco97}. A total of 10 bias fields were generated for the simulated dataset. The bias simulation settings include a sampling half-width of $L=12$ per dimension and sampling rate of $(0.5,1.75,1.75)$ in the $(\omega_z,\omega_y,\omega_x)$ Fourier space. For the sampling grid $G$, plane $\omega_z = 0$ was omitted to produce more pronounced bias fields.

For real-world data, MR patient scans were acquired on a Siemens 3T Vida with thorax and pelvis body regions being scanned using a 32 channel posterior spine array. All scans were acquired using a gradient-echo based VIBE Dixon dual echo sequence with lowest possible repetition time (TR) and time to echo (TE) values. Image resolution was $2\times 2\times 2\,$mm isotropic and patient images were retrospectively included in this Insitutional Review Board (IRB) approved study.

\section{Results and Discussion}

\begin{table*}
\centering
{\small
\begin{tabular}{ cc|c|c|c|cc} 
\hline
Experiment & Method &  Bias & Debias   & \multicolumn{2}{c}{Runtime [s]}
\\
& &  RMSE [1] & RMSE [1]  & CPU &  GPU \\
\specialrule{.13em}{0em}{0em} 
Line32 & N4 & 0.1684 & 2334 & 623 & --- \\\hline
Line32 & LapGM 1-seq. & 0.1889 & 2564 & 64.1 & 7.96  \\\hline
Line32 & LapGM 3-seq. & 0.0614 & 934.7 & 86.2 & 13.6  \\

\specialrule{.13em}{0em}{0em} 
Rect3\_Angn90 & N4 & 0.0738 & 2791 & 605 & ---  \\\hline
Rect3\_Angn90 & LapGM 1-seq. & 0.0720 & 2680 & 53.7 & 7.78  \\\hline
Rect3\_Angn90 & LapGM 3-seq. & 0.0236 & 930.5 & 86.2 & 14.7  \\

\specialrule{.13em}{0em}{0em} 
Rect4 & N4 & 0.0729 & 2798  & 481 & --- \\\hline
Rect4 & LapGM 1-seq. & 0.0745 & 2624  & 84.3 & 7.76  \\\hline
Rect4 & LapGM 3-seq. & 0.0220 & 840.7 & 90.4 & 15.4  \\

\specialrule{.13em}{0em}{0em} 
Rect5 & N4 & 0.0699 & 2975  & 546 & ---  \\\hline
Rect5 & LapGM 1-seq. & 0.0687 & 2586 & 87.4 & 6.53  \\\hline
Rect5 & LapGM 3-seq. & 0.0607 & 1166 & 104 & 13.7  \\

\specialrule{.13em}{0em}{0em} 
Rect7 & N4 & 0.0566 & 2508 & 600 & ---  \\\hline
Rect7 & LapGM 1-seq. & 0.0787 & 3223 & 71.1 & 7.99  \\\hline
Rect7 & LapGM 3-seq. & 0.0245 & 1046 & 90.5 & 16.2  \\
\hline
\end{tabular}
}
\caption{Debias test results on BrainWeb augmented data.}\label{tab:1}
\end{table*}

Let $B^e, X^e$ be defined by the element-wise exponentiation
$$B^e_i\coloneqq\exp(B)\quad\text{and}\quad X^e_i \coloneqq \exp(X_i).$$
Simulated data were evaluated using the following metrics:
\begin{align*}
\text{Bias RMSE} &= \bigg(\frac{1}{|\Omega|}\sum_{i\in \Omega}\big(B^e_i - \hat{B}^e_i\big)^2\bigg)^{1/2},\\
\text{Debias RMSE} &= \bigg(\frac{1}{|\Omega|}\sum_{i\in \Omega}\big(X_i^e/B_i^e - X_i^e/\hat{B}_i^e\big)^2\bigg)^{1/2}.
\end{align*}
For notation $\hat{B}$ is the estimated bias field are $\Omega$ is the set of indices contained within the tissue mask provided by BrainWeb. Simulated test results for N4 and LapGM can be seen in Table \ref{tab:1}.

\subsection{Simulated Data}

\begin{figure}[t]
\centering
\includegraphics[width=8.75cm]{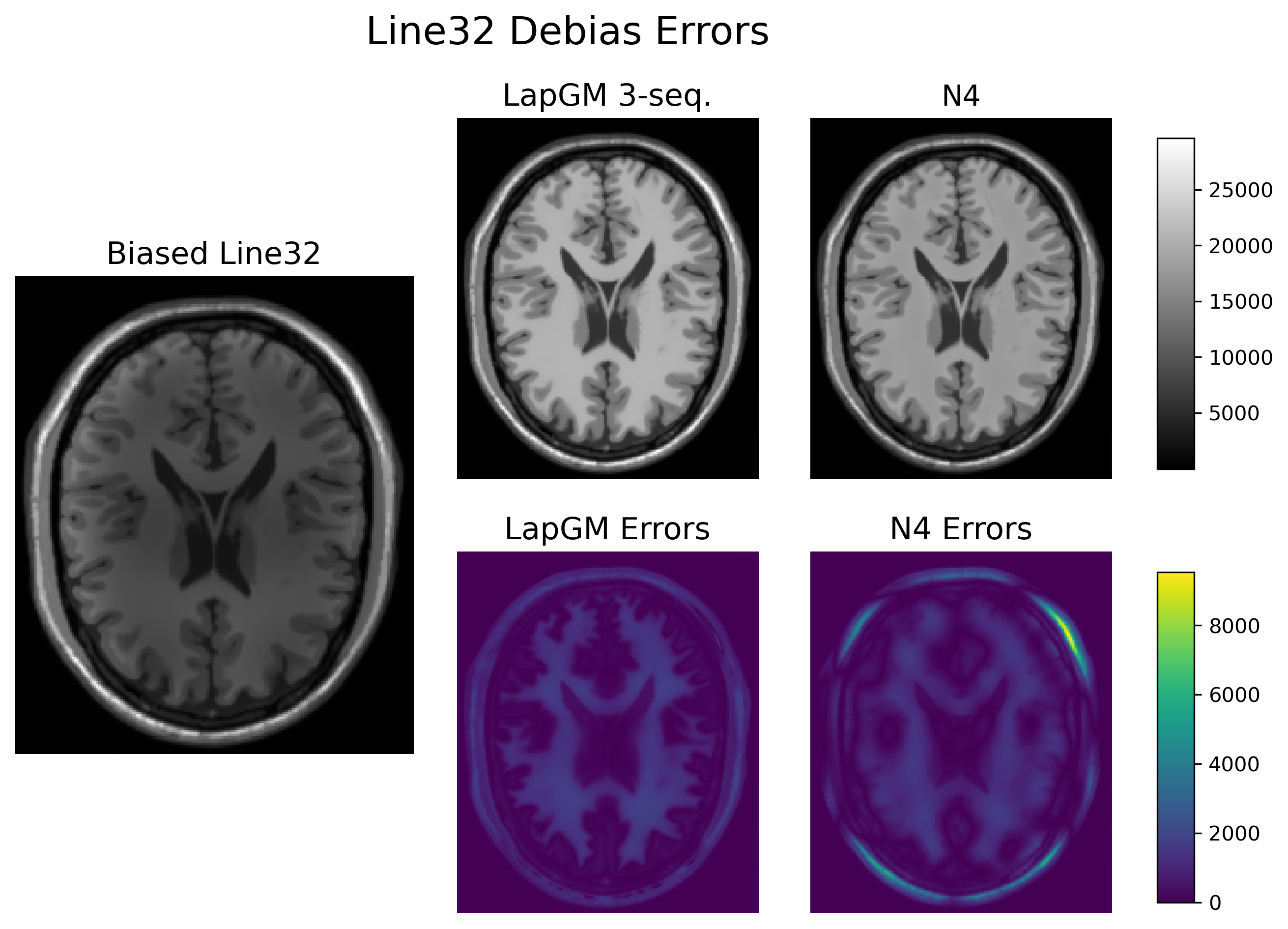}
\caption{Debias error results on Line32 with N4 and LapGM 3-seq. algorithm. Second column of this figure reveals LapGM's tendency to increase contrast for certain dominant tissue groups.}\label{fig:l32}
\end{figure}

\begin{figure}
\centering
\includegraphics[width=8.75cm]{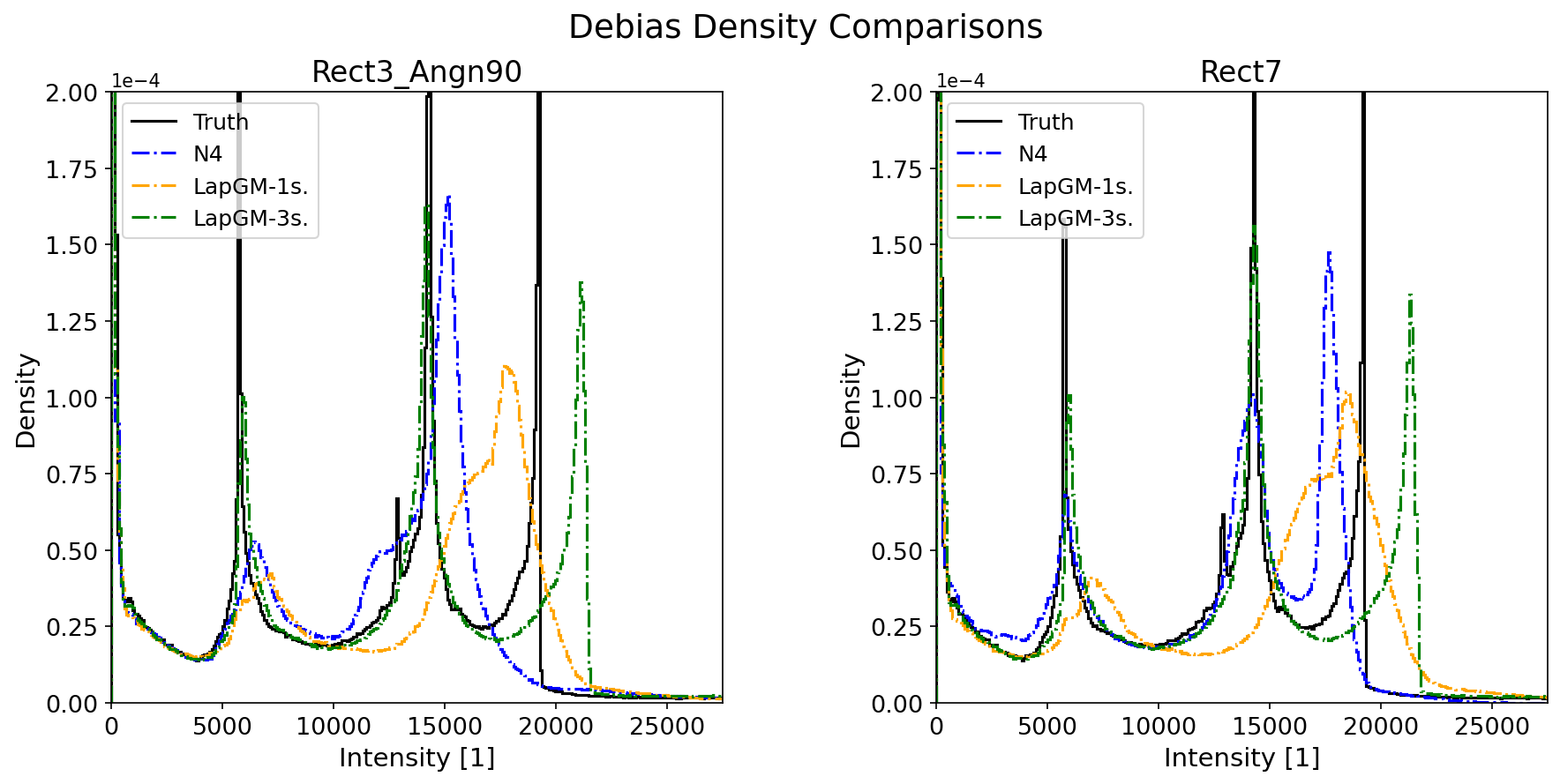}
\caption{Probability density comparison between true tissue density and debiased tissue density functions.}\label{fig:bias_hists}
\end{figure}

Table \ref{tab:1} shows comparable performance between N4 and LapGM in the single-sequence setting and superior LapGM performance in the multi-sequence case. Large consistent improvements can be found in the debias RMSE for LapGM 3-sequence. The metric with the smallest improvement was tissue total variation. Fig. \ref{fig:l32} and \ref{fig:bias_hists} have been provided to better understand performance differences in these metrics. Table \ref{tab:1} also shows that, for the given testing environment, LapGM runs significantly faster that N4 with a near 80-fold improvement in runtime for the GPU-accelerated case.

Fig. \ref{fig:l32} shows the debiasing result for the Line32 biased experiment. N4 and LapGM show similar errors in the central tissue group of the BrainWeb phantom, with LapGM's errors closely following the contour of the central tissue group. At the boundary of the phantom's anatomy we see a clear difference between the errors of N4 and LapGM. Here N4 has trouble adapting to the spatial variations found in the simulated bias field. As mentioned before, multi-channel RF configurations have the tendency to sharply increase bias at the boundary of the patient's anatomy. This is something we are able to account for when setting up the graph Laplacian for LapGM.

Fig. \ref{fig:bias_hists} shows the recovered tissue distributions for N4 and LapGM. A few features we will focus on in the recovered tissue distributions are: number of peaks recovered, width/location of peaks, and recovery consistency. With these qualitative metrics we will be able to glean more insight into how the N4 and LapGM methods work.

First is the number of peaks recovered. In this regard LapGM 3-seq. is able to consistently identify major tissue groups between differently biased examples. Depending on the kind of bias field, a correctly calibrated N4 may sometimes recover the major tissue groups. In the 1-sequence setting, LapGM does not have as much data to contrast different tissue groups. As such, LapGM 1-seq. shows the tendency of overlapping certain tissue groups which are similar in intensity.

For the width and location of peaks, LapGM 3-seq. shows the sharp recovery with each peak being near its original location. N4's performance depends on the number of peaks recovered but in general shows a peaked recovery. Here LapGM 1-seq. is faced with the same issue from before with peak width and location being off for the last two tissue groups.

Next is the topic of consistent recovery. To LapGM 1-seq.'s benefit this is a category it performs fairly well in. LapGM 3-seq. performs similarly well and N4 has issues with consistent distribution recovery. Inconsistent distribution recovery has negative downstream impacts to any supervised learning model. Modern machine learning methods can be remarkably good at accounting for missing or corrupted information, as long as this missingness or corruption is consistent between data samples. In the case of N4, inconsistent distribution recovery could lead to contradictory training signals for supervised methods which rely on N4 for data cleaning.

Lastly we identify the outlier on LapGM 3-seq. bias RMSE performance on testing data Rect5. Rect5 features five rectangular coils in a pentagon pattern arround the BrainWeb phantom. The deviation in LapGM 3-seq.'s perfomance may be an issue of incomplete convergence during the optimization process. In practice, incomplete optimization can be evaluated by analyzing whether the final bias field estimate of LapGM is too smooth or too rough.  After identifying incomplete optimization can be corrected for by modifying the regularization strength $\tau^{-1}$ for the example in question.

\subsection{Patient Data}

\begin{figure}
\centering
\includegraphics[width=8.75cm]{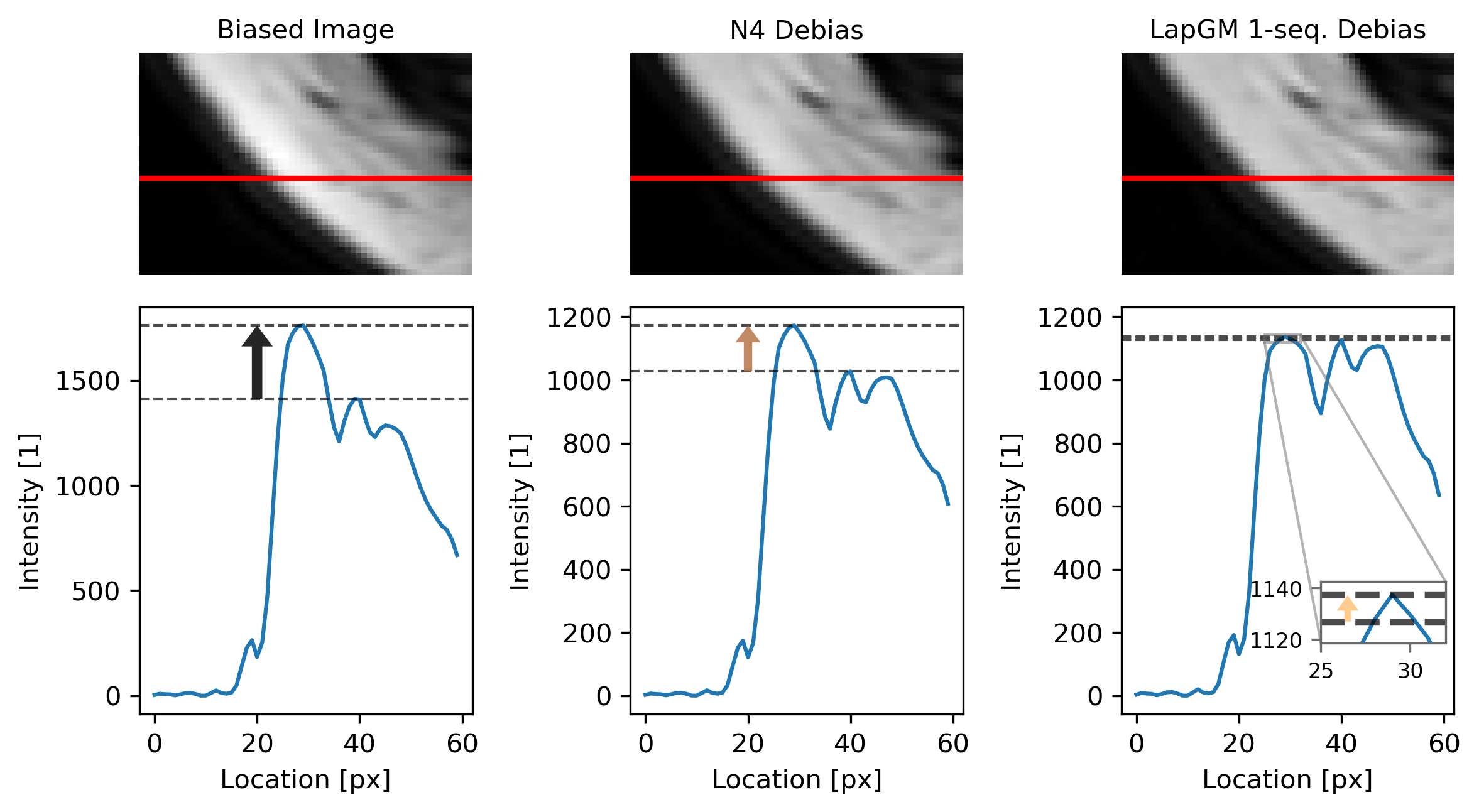}
\caption{Line profiles of the lower-left section of patient pelvic region. Line profile in red (top) with corresponding intensity shown in blue (bottom). Top dotted line is placed on dotted peak and the bottome dotted line is placed on the second peak next to the trough.}\label{fig:line_prof}
\end{figure}

\begin{figure*}
\centering
\includegraphics[width=17.5cm]{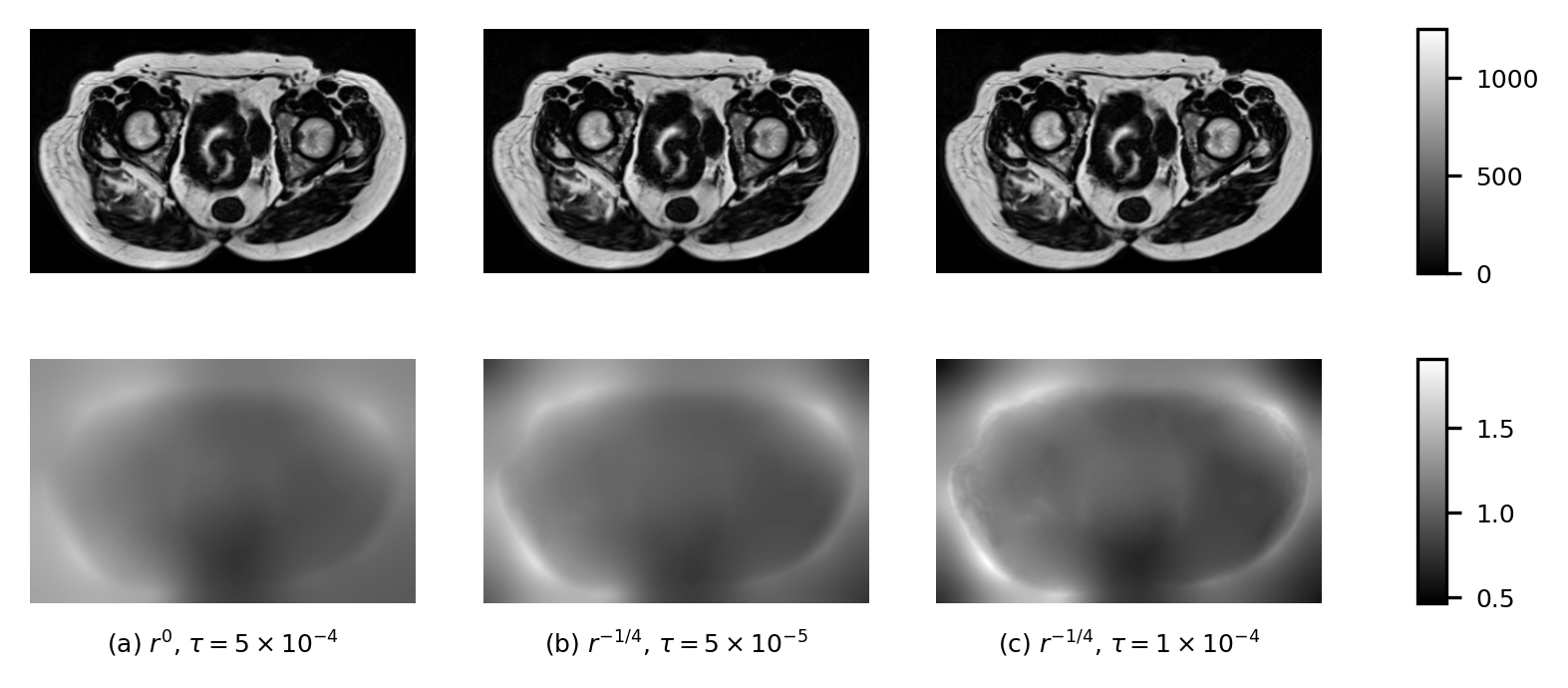}
\caption{Debias and bias field comparison of single sequence LapGM for varying regularizations $\tau$ and gradient penalty weights $h_\alpha(r,z)$.}\label{fig:bias_comp}
\end{figure*}

For our patient scan analysis we will focus on the debiasing results of a patient's pelvic between N4 and the single sequence LapGM model, in particular we will be focussing on how image contrast can be balanced for improved bias correcting performance. Fig. \ref{fig:line_prof} shows slice line profiles for the lower-left section of a patient scan. The full slice with its bias comparison can be found in Fig. \ref{fig:bias_comp}. The leftmost column of Fig. \ref{fig:line_prof} shows a compact but bright bias peak at around the 25 pixels mark. At this point, N4 shows a rough 60\% reduction in bias while the LapGM single-sequence model shows an almost complete removal of bias. The degree of bias correction for LapGM can be modified through the regularization stength parameter $\tau^{-1}$. Although stronger debiasing capabilities come at the cost of image contrast, this setting of the LapGM single-sequence model is able to preserve meaningful contrast. This can be verified by noting both N4 and LapGM share similar trough amplitudes for the 35 pixel mark.

Fig. \ref{fig:bias_comp} gives a visual comparison on trade-off between bias removal and image contrast for different $\tau$ regularizations and edge-weight settings. With decaying gradients weights, LapGM can fit for bias fields that are stonger along the extremity of the patient's anatomy. Fig. \ref{fig:bias_comp} also shows that the rate at which regularization $\tau$ trades contrast for bias correction may depend on the chosen edge-weight function $h_\alpha(r)$. As shown in sub-Fig. \ref{fig:bias_comp}b, practictioners may choose to leave a weak underlying bias field for better image contast.


\subsection{Normalization Comparisons}

\begin{figure}[b]
\centering
\includegraphics[width=8.75cm]{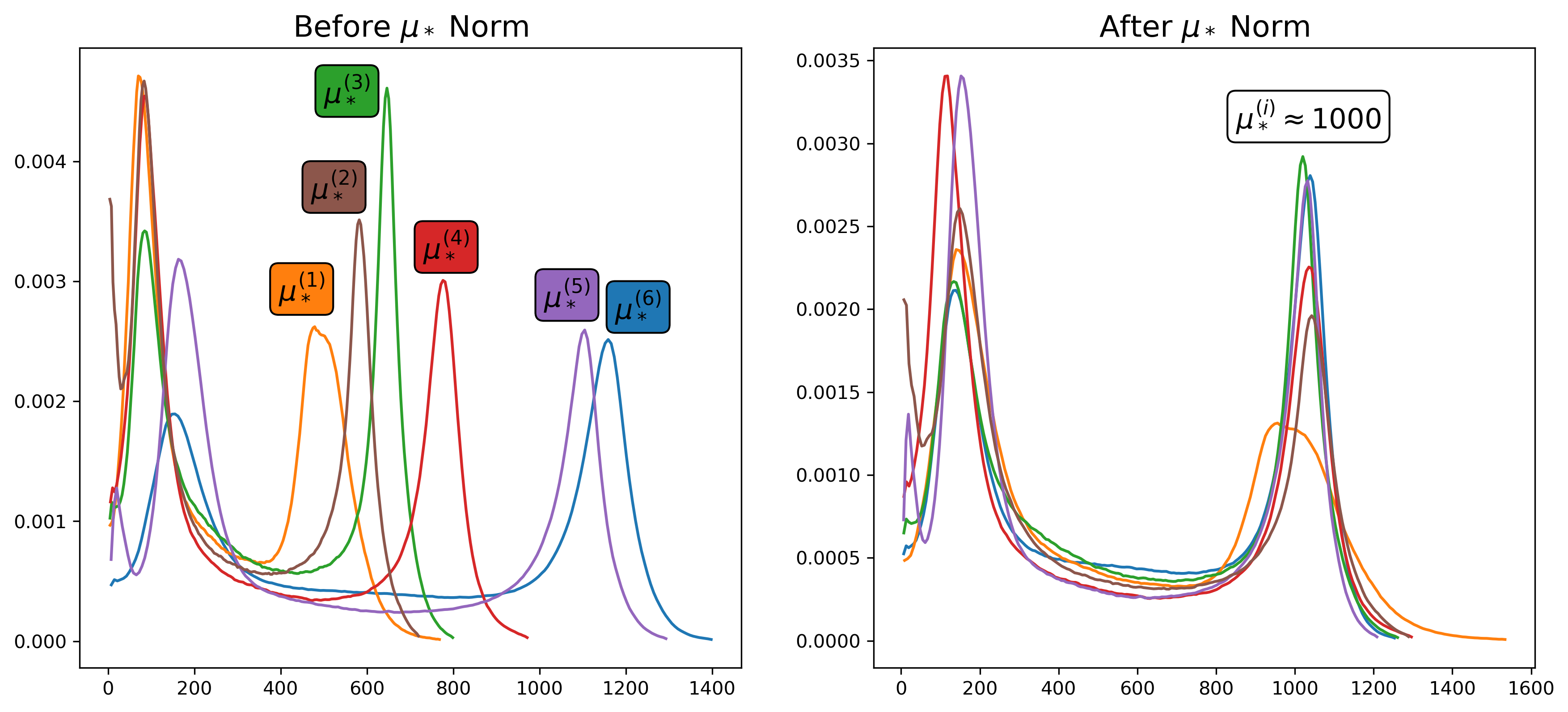}
\caption{Visualization of the $\mu_*$ normalization technique with real patient data. After normalization, the second peak of each patient distribution clusters around the target intensity of 1000.}\label{fig:mu_norm}
\end{figure}

We will refer to any normalization scheme that utilizes the fitted Gaussian parameters of LapGM as a LapGM-based normalization scheme. In its full generality, an involved LapGM-based normalization scheme could incorporate both class posterior probabilities $w$ with Gaussian information $\theta$. That said, it is still possible to get good normalization results with simple LapGM-based normalization schemes. In this section we analyze a LapGM-based normalization called $\mu_*$ normalization which is done by taking the largest fitted mean value
$$\mu_* = \max_{k\in[K]} \mu_k$$ 
and applying the appropriate scaling factor $\beta$ which scales $\mu_*$ to some target intensity value of choice. A visualization of the $\mu_*$ normalization process is shown in Fig. \ref{fig:mu_norm}.

\begin{figure*}
\centering
\includegraphics[width=17.5cm]{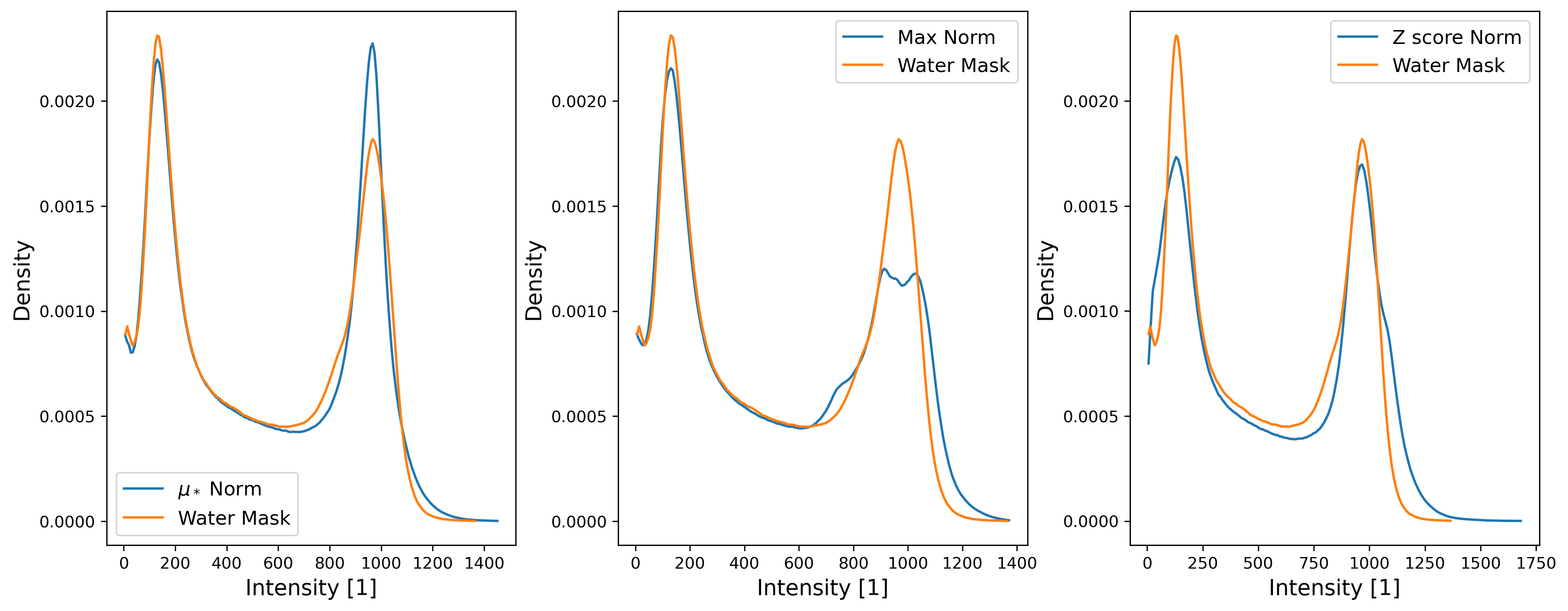}
\caption{Combined tissue distributions of 10 patient scans using different normalization techniques. The distributions produced from each normalization technique is peak-aligned with the tissue distribution of a water mask ROI normalized distribution. Starting from the leftmost column the tissue TV errors relative to the water mask ROI are 9.52\%, 14.0\%, and 21.7\%.}\label{fig:norm_comp}
\end{figure*}

The LapGM $\mu_*$ normalization was compared to the max normalization and Z-score normalization techniques. These comparisons were done relative to a water-masked region-of-interest (ROI) normalization baseline. The intensity distribution produced from each normalization is shown in Fig. \ref{fig:norm_comp}. A total of 10 LapGM-debiased patient scans were used for the normalization comparison. To allow for meaningful comparisons between the techniques, all three normalizations were peak-aligned to the baseline water mask ROI normalization.

Tissue total variations were computed for each of the peak-aligned distributions. The LapGM $\mu_*$ normalization had a TV error of 9.52\%, the max normalization had a TV error of 14.0\%, and the Z-score normalization had a TV error 21.7\%. The Z-score normalization featured a significant left tail which was not show in Fig. \ref{fig:norm_comp}. This left tail contributed Z-score's larger TV error calculation. 

One important qualitative difference between the recovered distributions of $\mu_*$ normalization and the other normalization techniques, is the presence of bumps along the recovered tissue distributions. Small bumps, like the ones visible along the max and Z-score normalization distributions, are an indication of intensity scaling mismatch between different patients. As shown by Fig. \ref{fig:mu_norm}, each of the 10 patient distributions feature a prominent second peak. Incorrect inter-patient scalings will cause these peaks to be misaligned and as a result form small bumps along the recovered tissue distribution. Note that this may even be the case for the small bump visible at the 800 intensity value for the water mask ROI distribution. As each water mask ROI is calculated by hand, it is possible for slight errors to accrue in the final aggregate tissue distribution.

This analysis was carried out on images where clear contours could be made for the different water ROIs, however we note that similar peak recovery can achieved by applying $\mu_*$ normalization to fat-dominant MR sequences as well.

\section{Conclusion}
In this paper, a multi-sequence debiasing algorithmic alternative to the common N4 debiasing algorithm is proposed. We have shown superior performance in various metrics for the multi-sequence case, as well as ease-of-use and interpretability in the single-sequence case. Both methods were tested on a variety simulated bias field configurations specified by \cite{Vinas22} as well as real-world patient data. Implementations for LapGM and the bias generation can be found in the Python packages \texttt{lapgm} and \texttt{biasgen} respectively.

In this paper a normalization technique using the fitted Gaussian parameters of LapGM is also proposed. This normalization is competitive with manual water mask ROI techniques and superior to max normalization and Z-score normalization techniques. Only one of the many possible LapGM-based normalization technique were analyzed. It may be possible to further improve normalization performance by utilizing more of LapGM's Gaussian parameters.

Future work could probe some theoretical properties of LapGM or attempt to explain some of the distributional tendencies of LapGM. As Fig. \ref{fig:bias_hists} showcases, N4 has a tendency to produce tissue peaks which vary based on bias field, while LapGM, both in the single and multi-sequence settings, seems to produce stable tissue distributions. Some other interesting phenomena to explore include why debias methods in general struggle to disentangle peak locations of high intensity tissue groups and under what conditions will multi-sequence LapGM increase tissue contrast rather than decrease it.

\appendices

\section{Expectation-Maximization for MAP}\label{app:A1}
Let $X \coloneqq \{X_{i}\}_{i=1}^n$. We are interested in the maximum a posteriori probability (MAP) for
$$\argmax_{\theta,B} p_\theta(B\,|\,X),$$
where $\theta$ and $B$ are conditionally independent from each other and an improper uniform prior $p(\theta) = 1$ is assumed on $\theta$. As the argmax is invariant to scalings and monotonic transforms, it is equivalent to consider the log joint probability
$$\argmax_{\theta,B} \log p_\theta(X,B),$$
where normalization factor $p_\theta(X)$ has been dropped.

A function $f(\psi)$ is said to be minorized by $g(\psi\,|\,\psi^{(t)})$ at $\psi = \psi^{(t)}$ if
$$g\big(\psi\,|\,\psi^{(t)}\big)\leq f(\psi),\;\forall\psi\quad\text{and}\quad g\big(\psi^{(t)}\,|\,\psi^{(t)}\big) = f\big(\psi^{(t)}\big).$$
For shorthand let $X \coloneqq \{X_i\}_{i=1}^n$. The goal will be to show 
\begin{align*}
g\big(\theta,B\,|\,\theta^{(t)},B^{(t)}\big) &\coloneqq Q\big(\theta,B\,|\,\theta^{(t)},B^{(t)}\big)\\
&\quad + \log p_{\theta^{(t)}}\big(X,B^{(t)}\big)\\
&\quad - Q\big(\theta^{(t)},B^{(t)}\,|\,\theta^{(t)},B^{(t)}\big)
\end{align*}
is a minorizing function of $(\theta,B)\mapsto \log p_\theta(X,B)$ at $(\theta^{(t)},B^{(t)})$ where
$$Q\big(\theta,B\,|\,\theta^{(t)},B^{(t)}\big) \coloneqq \mathbb{E}_{p_{\theta^{(t)}}(Z\,|\,X,B^{(t)})}[\log p_\theta(Z,X,B)].$$
This can be done by introducing an auxilliary distribution $q(Z)$ and decomposing the KL divergence of $q$ and $p_\theta$
$$D(q\,||\,p_\theta) \coloneqq \mathbb{E}_{Z\sim q}\Big[\log\frac{q(Z)}{p_\theta(Z\,|\,X,B)}\Big].$$
With some manipulation
\begin{align*}
D(q\,||\,p_\theta) &= \log p_\theta(X,B) + \mathbb{E}_q\big[\log q(Z) - \log p_\theta(Z,X,B)\big]\\
&= \log p_\theta(X,B) - \mathbb{E}_{q}[\log p_\theta(Z,X,B)] \\
&\quad  + \mathbb{E}_q\big[\log\big( q(Z)p_{\theta^{(t)}}(X,B^{(t)})\big)\big]\\ 
&\quad -\log p_{\theta^{(t)}}(X,B^{(t)}).
\end{align*}
Now set $q(Z) = p_{\theta^{(t)}}(Z\,|\,X,B^{(t)})$ and rearrange
\begin{align*}
\log p_\theta(X,B) &= Q\big(\theta,B\,|\,\theta^{(t)},B^{(t)}\big) + \log p_{\theta^{(t)}}\big(X,B^{(t)}\big)\\
&\quad - Q\big(\theta^{(t)},B^{(t)}\,|\,\theta^{(t)},B^{(t)}\big) + D(p_{\theta^{(t)}}\,||\,p_\theta).
\end{align*}
The non-negativity and equality properties of KL divergence
$$D(q\,||\,p)\geq 0,\; \forall q\quad \text{and}\quad D(q\,||\, p) = 0, \text{ if } p=q,$$
show that $g\big(\theta,B\,|\,\theta^{(t)},B^{(t)}\big)$ is indeed a minorizing function of $(\theta,B)\mapsto \log p_\theta(X,B)$ at $(\theta^{(t)},B^{(t)})$. Lastly since $g\big(\theta,B\,|\,\theta^{(t)},B^{(t)}\big)$ only has one term which depends on $(\theta$, $B)$, the maximization step may be simplified to
\begin{align*}
\big(\theta^{(t+1)},B^{(t+1)}\big) &= \argmax_{\theta, B} g\big(\theta,B\,|\,\theta^{(t)},B^{(t)}\big)\\
&= \argmax_{\theta, B} Q\big(\theta,B\,|\,\theta^{(t)},B^{(t)}\big).
\end{align*}

\section{MAP Coordinate Updates}\label{app:A2}
With the shorthands $a_{ik} \coloneqq X_i - \mu_k - B_i \mathbbm{1}_m$ and $w_{ik} = p_{\theta^{(t)}}\big(Z_i = k\,|\,X_i, B_i^{(t)}\big)$, we expand the following objective
\begin{align*}
    -Q\big(\theta,B\,|\,\theta^{(t)},B^{(t)}\big) &\propto \sum_{i=1}^n\sum_{k=1}^K w_{ik}\Big( \frac{1}{2}a_{ik}^\top \Sigma_k^{-1} a_{ik}  - \frac{1}{2}\log |\Sigma_k^{-1}|\\  &\quad-\log\pi_k\Big)+ \frac{1}{2\tau}B^\top L B.
\end{align*}
We will be interested in the optimization
$$\min_{\substack{\mu\in\mathbb{R}^m,\, B\in\mathbb{R}^n,\\ \Sigma_k\in\mathbb{R}^{m\times m}:\;\Sigma_k\succ 0,\\ \pi_{k}\in[0,1]:\;\sum_{k=1}^{K}\pi_k = 1}} -Q\big(\theta,B\,|\,\theta^{(t)},B^{(t)}\big). $$
Parameter $\pi_k$ can be independently optimized as
$$\pi_k^* = \frac{1}{n}\sum_{i=1}^n w_{ik}.$$
The simplified objective
$$\mathcal{L} = \sum_{i=1}^n\sum_{k=1}^K \frac{w_{ik}}{2}\Big( a_{ik}^\top \Sigma_k^{-1} a_{ik}  - \log |\Sigma_k^{-1}|\Big)+ \frac{1}{2\tau}B^\top L B$$
can be optimized as an unconstrained problem
\begin{align*}
    \min_{\substack{\mu\in\mathbb{R}^m,\, B\in\mathbb{R}^n\\ \Sigma_{k}^{-1}\in\mathbb{R}^{m\times m}:\,\Sigma_k^{-1}\succ 0}}\mathcal{L}(\mu, \Sigma^{-1}, B),
\end{align*}
where for $\Sigma_k\succ 0$ it is equivalent to optimize over the reparametrized $\Sigma_k^{-1}$. The local optimality conditions of $\mathcal{L}$ are
\begin{align*}
&\frac{\partial\mathcal{L}}{\partial\mu_k} =\sum_{i=1}^n -w_{ik}\,\Sigma_k^{-1}a_{ik} = 0,\\
&\frac{\partial\mathcal{L}}{\partial\Sigma_k^{-1}} = \sum_{i=1}^n w_{ik}\, a_{ik}a_{ik}^\top - w_{i k}\Sigma_k = 0,\\
&\frac{\partial \mathcal{L}}{\partial B_i} = \sum_{k=1}^K -w_{ik}\mathbbm{1}_m^\top \Sigma_k^{-1} a_{ik} +\frac{1}{\tau}(LB)_i = 0.
\end{align*}
This last condition can be rewritten in vector notation
$$\frac{\partial \mathcal{L}}{\partial B} =\sum_{k=1}^K \text{diag}(w_{\cdot k})\, (X - \mu_k\mathbbm{1}_n - \mathbbm{1}_m B^\top)^\top\Sigma_k^{-1} \mathbbm{1}_m -\frac{1}{\tau}LB,$$
where $\text{diag}(w_{\cdot k})$ is a diagonal matrix with entries $w_{ik}$. Parameter updates can be done in a block-cooridinate fashion until convergence.

After sufficiently optimizing to a new set of parameters $(\theta',B')$, $w_{ik}$ can be updated as
$$w_{ik} = \frac{\pi'_k\, p_{\theta'}\big(X_i\,|\,Z_i=k,B'\big)}{\sum_{\ell=1}^K\pi'_l\, p_{\theta'}\big(X_i\,|\,Z_i=\ell,B'\big)}.$$ 
For practical applications, the authors have found that sequential updates of the form $(w\rightarrow\theta\rightarrow B\rightarrow w\rightarrow\ldots)$ yield quick and stable convergences.

As a final comment, we note that neither the original log-likelihood nor its expectation-step surrogate is necessarily convex for all $\Sigma_k^{-1} \succ 0$. For the case of a standard Gaussian mixture, it has been shown \cite{Chi16} that overall performance is dependent on the quality of initial estimate parameter. In practice, a K-means initialization step with a potential data transform can help to avoid bad local minima.

\printbibliography

\end{document}